\documentclass[floatfix,10pt,onecolumn,showpacs,preprintnumbers,amsmath,amssymb]{revtex4}

\usepackage{epsf}
\usepackage{graphicx}  
\usepackage{dcolumn}   
\usepackage{bm}        

\newcommand{\be}{\begin{equation}}
\newcommand{\en}{\end{equation}}
 \newcommand{\bea}{\begin{eqnarray}}
 \newcommand{\ena}{\end{eqnarray}}
  \newcommand{\van}{van der Waals-Maxwell}

\begin{document}

\title{Critical behaviors of gravity under quantum perturbations}
\author{Hongsheng Zhang\footnote{Electronic address: hongsheng@shnu.edu.cn} }
\author{Xin-Zhou Li \footnote{Electronic address: kychz@shnu.edu.cn} }
 \affiliation{ Center for
Astrophysics , Shanghai Normal University, 100 Guilin Road,
Shanghai 200234,China}

\date{ \today}

\begin{abstract}
 Phase transition and critical phenomenon is a very interesting topic in thermodynamics and statistical mechanics. Gravity is believed to has deep and inherent relation to thermodynamics. Near the critical point, the perturbation becomes significant. Thus for ordinary matter (govern by interactions besides gravity) the critical behavior will become very different if we ignore the perturbations around the critical point, such as mean field theory. We find that the critical exponents for RN-AdS spacetime keeps the same values even we consider the full quantum perturbations. This indicates a key difference between gravity and ordinary thermodynamic system.

\end{abstract}


\preprint{arXiv: }
 \maketitle

\section{Introduction}
   The relation between gravity theory and thermodynamics is an interesting and
   profound issue. The key quantities bridging the gravity and thermodynamics are
   temperature and entropy.  Temperature of an ordinary system denotes the average kinetic energy
   of microscopic motions of a single particle. To gravity the temperature becomes subtle. Since we do not
   have a complete quantum theory of gravity, for general case we can not use the
   usual way to get the temperature of the gravity field. Under this situation one can
   set up some thermodynamics and statistical quantities of gravity by
   using semi-quantum (matter field is quantized, but gravity remains classical)
   theory, though the concept of gravitational particle is not clear.  The black hole
   thermodynamics (in fact, spacetime thermodynamics, because  the physical quantities in black hole thermodynamics
    should be treated as the quantities of the globally
   asymptomatic flat manifold) is set up in \cite{4law} and confirmed by Hawking radiation \cite{hawk}.

   The Hawking radiation is
   a pivotal discovery in the history of black hole thermodynamics, which has been confirmed in several different ways. An intuitive reasoning of
   Hawking radiation in tunneling scenario is proposed in \cite{parikh}. This method not only realized some early intuitive thought on the Hawking radiation, but find the correction to the
   blackbody spectrum of the radiation. The correction to the original Hawking radiation is easy to understand: In the early work, the mass of black hole is assumed to be a constant, which is of course only an approximation since energy flows away with the radiations. In the original tunneling scenario, semi-classical approximations are essential in the deduction. The full quantum perturbations, i.e., the quantum corrections to all orders for Hawking radiation, is found in \cite{banerjee}. More importantly, this study of quantum perturbations is a self-consistent approach. The corrections for entropy, temperature and other quantities are given in the same frame, in which the first law of thermodynamics is satisfied. The leading correction for entropy just takes a log form, which has been obtained several times in several studies from many aspect of black hole entropy. It can be obtained from thermal perturbations, quantum perturbations, loop quantum gravity, Euclidean path integral approach, etc \cite{page}. Most of the early approaches are not self-consistent: Only the correction to entropy is considered, which breaks the first law. Consistent black hole thermodynamics with perturbation effects from very different aspects can be found in \cite{cai1}.

   More and more analogies are found between ordinary thermodynamics and black hole thermodynamics through more and deep studies. The four laws of black hole dynamics can be mapped to the four law of thermodynamics. Every quantity finds its counterpart in ordinary thermodynamics. It is generally known that phase transition and critical phenomenon play a central role in modern thermodynamics and statistical mechanics. The phase transition is also discovered in black hole physics. To describe phase transition and critical phenomenon by using an electrodynamic analogy has a fairly long history \cite{TDL}.  In an asymptotic AdS space, a first order phase transition occurs in a charged black hole (Hawking-Page phase transition) \cite{HP}. AdS gets increasing attentions since the AdS/CFT correspondence was proposed\cite{adscft}. A most simple version of AdS/CFT says that gravity in AdS space exactly maps to a supersymmetric Yang-Mills theory on its boundary. This can be treated as a perfect example of holographic principle. The Hawking-Page phase transition maps to quark confinement/definement transition in the CFT part \cite{witt}, such that gets more physical significance. Moreover, RN-AdS black undergoes a second order phase transition and a critical point appears, which is very similar to what happens in \van~ gas-liquid system \cite{em}. The behavior of the electric potential and the charge play the role of volume and pressure in \van ~ gas-liquid system. However, this is only an apparent analogy from the plots of electric potential vs. charge and volume vs. pressure. The critical exponents are essentially different between the two systems.

    The \van~ gas-liquid is a typical instance of mean field theory (MFT). MFT was suggested many times in history, including Weiss molecular field theory, Landau-Ginzburg superconductor model, Bragg-Williams approximation, etc. All of them share the same critical exponents. The essence of MFT is to ignore the perturbations around the critical point: it treats all effects  working on a particle from all the other particles in a system as a ``Mean Field". When the phase transition in RN-AdS space time is explored, no perturbation is considered. The other important property of the phase transition in RN-AdS is the critical exponents are dimension-independent \cite{wuxiao2}. This is also a key property of MFT. We may habitually conjecture that the critical exponents of RN-AdS take the same values of MFT. But the result is surprising in some sense. It is different from neither  MFT nor Kadanoff-Wilson (renormolization group) theory. However it inherits an exponent from MFT ($\delta$) and an exponent from renormalization group ($\beta$), and has its own exponents ($\alpha$, $\gamma$).  The spirit of Kadanoff-Wilson theory is that there is no length scale in a system at the critical point. But, it is an inherent scale in gravity theory $G^{-1/2}$, which generates the difficulties to construct a no-scale theory for gravity at the critical points. However, we can include the effects of the perturbations in a very different way. In this article we develop a new method to consider the full quantum perturbations at the critical points.

    This article is organized as follows. In the next section we review some properties of RN-AdS and study its critical exponents with quantum perturbations. In section 3 we present our conclusion.

    \section{Critical exponents of RN-AdS with full quantum perturbations}
  The thermodynamics of RN-AdS in different dimensions have discussed by many authors by several different methods \cite{many}. First, we revisit the principle results of 4-dimensional RN black hole in asymptotic AdS space. The metric reads,
  \begin{equation} \label{e1}
ds^2  =  - f(r)dt^2  + f(r)^{ - 1} dr^2  + r^2 d\Omega ^2,
\end{equation}
where $d\Omega ^2$ denotes a unit 2-sphere and
\begin{equation}
\label{e24}
f(r) = 1 - \frac{2M }{r} + \frac{{q^2 }}{{r^2 }} + \frac{{r^2
}}{{l^2 }}
\end{equation}
$M$ and $q$ label the ADM mass associated to the Killing vector $\frac{\partial}{\partial t}$ and electric charge of the black
hole respectively (We adopt nature unit $ G = c = \hbar  = k_B = 1 $ without notations). Generally RN-AdS black hole has two horizons,
inner Cauchy horizon and outer event horizon, located at $r_-$ at
$r_+$ respectively. The two horizons are determined by mass, electric charge, and the cosmological constant,
\begin{equation}
2M = r_ +   + r_ -   + \frac{{r_ +  ^4  - r_ -  ^4 }}{{l^2 (r_ + -
r_ -  )}}
\label{mass}
\end{equation}
\begin{equation}
q^2  = r_ +  r_ -  \left(1 + \frac{{r_ +  ^3  - r_ -  ^3 }}{{l^2
(r_ + - r_ -  )}}\right).
\label{charge}
\end{equation}
 When $q^2  \ge M ^2 $ a naked
singularity will appear. As usual, the black hole entropy $S$ is a quarter of the outer horizon,
\begin{equation}
S = \frac{1}{4}A = \pi r_ +  ^2.
\label{entropy}
\end{equation}
The first law $dM=TdS +\phi dq$ yields the temperature and the electric potential,
\begin{equation}
\label{temp}
T = \left( {\frac{{\partial M }}{{\partial S}}} \right)_q  =
\frac{{(r_ +   - r_ -  )(l^2  + 3r_ + ^2  + 2r_ +  r_ -   + r_ -
^2 )}}{{4\pi l^2 r_ + ^2 }}
\end{equation}
and
\begin{equation}
\phi  = \left( {\frac{{\partial M }}{{\partial q}}} \right)_S  =
\frac{{\sqrt {r_ +  r_ -  [1 + (r_ + ^2  + r_ +  r_ -   + r_ - ^2
)/l^2 ]} }}{{r_ +  }} = \frac{{q}}{{r_ +  }}.
 \label{potential}
\end{equation}
The EOS of the RN-AdS black hole $f(T,q,\phi)=0$ can be obtained by
using Eq.~(\ref{mass})-Eq.~(\ref{potential}),
 \be
4 \pi Tq\phi={\phi^2-\phi^4+\lambda q^2},
 \label{tphiq}
  \en
  where ${\lambda}/{3}=1/l^2$.
 The critical values of temperature $T_c$, charge $q_c$, and potential $\phi_c$ can be calculated routinely,
 \be
 T_c=\frac{\sqrt{2\lambda}}{3 \pi },
 \en
 \be
 q_c=\frac{1}{\sqrt{12\lambda}},
 \en
 \be
 \phi_c=\frac{1}{\sqrt{6}}.
 \en
 The critical exponents around the critical points are defined as follows.
 \begin{eqnarray}
\nonumber
 && (1)\quad q - q_c  \sim \left| {\phi  - \phi _c } \right|^\delta
\quad (T = T_c ),
\\ \nonumber
&& (2)\quad \phi  - \phi _c  \sim \left| {T - T_c } \right|^\beta
\quad (q = q_c ),
\\ \nonumber
&& (3)\quad C_q  \sim \left| {T - T_c } \right|^{ - \alpha } \quad
(q = q_c ),
\\
&& (4)\quad \kappa _T  \sim \left| {T - T_c } \right|^{ - \gamma }
\quad (q = q_c ).
\label{criexp}
 \end{eqnarray}
  These exponents have been worked out \cite{wuxiao2},
  \be
  \delta=3,~~~\beta=1/3,~~~ \alpha=2/3,~~~~\gamma =2/3.
  \en
 Interestingly, these exponents are dimension-free, which is a significant property of MFT. It is not surprised since we do not include any perturbations around the critical point, where perturbations, in fact, dominates the whole process approaching the point. To include the perturbations will shift the physical quantities, such as temperature, pressure, entropy, etc.

 The quantum perturbation on a classical gravity background has been explored for several years. However, most of them are not self-consistent approaches in sense of thermodynamics. The corrected quantities do not satisfy the laws of thermodynamics, not like what classical theory can do. In principle, any from of  matters should obey thermodynamics, since thermodynamics does not depends on the concrete form of matter. A recent approach of quantum perturbation of black hole realizes a consistent thermodynamics after quantum corrections. In this approach,
 the corrected temperature $T_{qu}$ reads,

 \be
 T_{qu}=T\Big(1+\sum_i\beta_i\frac{\hbar^i}{M^{2i}}\Big)^{-1},
 \en
 where $\beta_i$ are dimensionless constants, $T$ represents the ``classical" Hawking temperature, and $M$ denotes the mass of the black hole. In the above equation we restore the Planck constant $\hbar$ to clear the order of the corrections. Thus, it is clear that in the classical limit ($\hbar\to 0$) the temperature degenerates to the original Hawking temperature. The series in bracket can be worked out if $(\beta_i\hbar)^i=k$,
 \be
 T_{qu}=T\Big(1-\frac{k}{M^{2}}\Big).
 \en
 This equation includes the total effects of the quantum perturbations to all orders. With a corrected entropy, the consistent first law can be obtained in this frame \cite{banerjee}. Note that there is no extra corrections to mass, charge and angular momentum. Thus, with quantum corrections the equation of state (\ref{tphiq}) becomes
 \be
 T_{qu}=\Big(1-\frac{k}{M^{2}}\Big)\left({\phi^2-\phi^4+\lambda q^2}\right),
 \label{qutphiq}
  \en
 where the mass $M$ relates to $q$ and $\phi$ by,
 \be
  M=\frac{1}{2} \left(\frac{q}{\phi}+\phi q+\frac{\lambda q^3}{3 \phi^3}\right).
  \en
  The second order phase transition occurs at
  \be
  \left(\frac{\partial q}{\partial \phi}\right)_T=0,
  \label{diff}
  \en
  and
  \be
  \left(\frac{\partial^2 q}{\partial \phi^2}\right)_T=0.
   \label{ddiff}
   \en
 (\ref{diff}) yields

 \be
  \frac{q \left\{q^2 \left(-\phi^2+3 \phi^4+\lambda q^2\right) \left(3 \phi^2+3 \phi^4+\lambda q^2\right)^3-36 k \phi^6 \left[-6 \phi^6+15 \phi^8+14 \lambda \phi^2 q^2+7 \lambda^2 q^4+\phi^4 \left(3-6 \lambda q^2\right)\right]\right\}}{\phi q^2 \left(-\phi^2+\phi^4+\lambda q^2\right) \left(3 \phi^2+3 \phi^4+\lambda q^2\right)^3+36 k \phi^7 \left[3 \phi^8-14 \lambda \phi^2 q^2-7 \lambda^2 q^4-\phi^4 \left(3+4 \lambda q^2\right)\right]}=0.
  \label{diffcom}
 \en
  It is an algebraic equation of degree 16 with respect to $\phi$, or degree 10 with respect $q$. Further, (\ref{ddiff}) yields an even more complicate algebraic equation. It is no hope to get any analytic solution. From the experiences of the previous studies of the critical points of second order phase transition, we develop a special method to get the critical point. In fact, we need not to get all the solutions of (\ref{diff}) and (\ref{ddiff}). The essential condition is that two roots of (\ref{diff}) degenerates to one. Under this condition, (\ref{diffcom}) with respect to $\phi^2\equiv x$ is equivalent to
  \be
  (x-x_c)^2(c_0+...+c_5x^5+x^6)=0,
    \en
    where $x_c$, $c_0,~..., c_5$ are constants.
  A comparison of the coefficients with (\ref{diffcom}) generates the following equation set,
   \be
   {c_0} x_c^2-\frac{\lambda^4 q^8}{81}=0,
   \en
   \be
   (-2 c_0 xc + c_1x_c^2) - \frac{8}{81} \lambda^3 q^6=0,
   \en
   \be
   \left({c_0}-2 {c_1} x_c+{c_2} x_c^2\right)-\frac{\left(18 {\lambda}^2 q^6+12 {\lambda}^3 q^8\right)}{81 q^2}=0,
   \en
   \be
   (c_1 - 2 c_2x_c + c_3x_c^2) - \frac{(180 k \lambda^2 q^4 + 72 \lambda^2 q^6)}{(81 q^2)}=0,
   \en
   \be
   \left({c_2}-2 {c_3} x_c+{c_4} x_c^2\right)-\frac{\left(-27 q^2+360 k {\lambda} q^2+108 {\lambda} q^4+54 {\lambda}^2 q^6\right)}{81 q^2}=0,
   \en
   \be
   (c_3 - 2 c_4x_c + c_5x_c^2) -\frac{ (324 k - 648 k \lambda q^2 + 216 \lambda q^4)}{
 81 q^2}=0,
 \en
   \be
   \left({c_4}-2 {c_5} x_c+x_c^2\right)-\frac{\left(-648 k+162 q^2+108 {\lambda} q^4\right)}{81 q^2}=0,
   \en
   \be
   (c_5 - 2x_c) -\frac{(-108 k + 216 q^2)}{81 q^2}=0.
   \en
  In the above set, we treat $x_c,q,~c_0,~c_1...,~c_5$ as variables. We do get the analytic solution of the above equation set. But, unfortunately, it is rather lengthy (more than ten pages for a single solution of one variable) and has no illumination in physical sense. Hence we do not display it here. We only show some numerical results to compare with the original RN-AdS case.
  \begin{figure}
\includegraphics{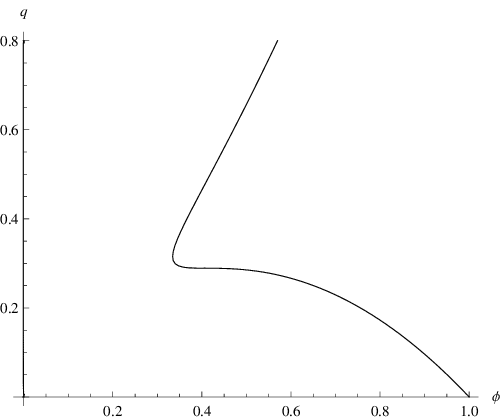}
\caption{The critical isotherm of RN-AdS black hole, in which $T_c=\frac{\sqrt{2}}{3 \pi }$, $\phi_c=\frac{1}{\sqrt{6}}$, $q_c=\frac{1}{2 \sqrt{3}}$.}
\label{rnads}
\end{figure}
 \begin{figure}
\includegraphics{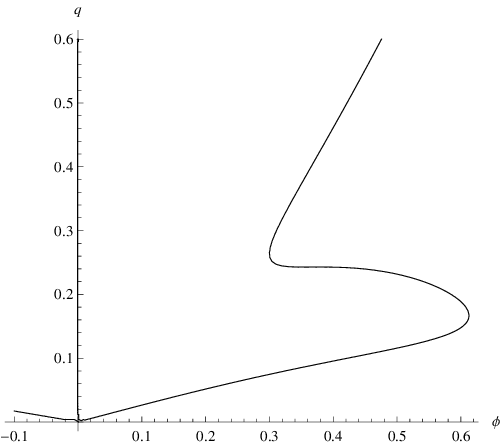}
\caption{The critical isotherm of perturbed RN-AdS black hole, in which we set $k=0.01, ~T_c=0.1479$, $\phi_c=0.3697$, $q_c=0.2435$.}
\label{k0.01}
\end{figure}

Fig \ref{rnads} displays the critical isotherm of RN-AdS black hole without perturbation, while Fig \ref{k0.01} illustrates the corresponding with a quantum perturbation. One sees that the shapes of the two figures become different when $q$ is small. However, the critical exponents depend only on the properties of the system around the critical point. By definition (\ref{criexp}), we find the critical exponents of the perturbed RN-AdS by series expansion,
   \be
  \delta=3,~~~\beta=1/3,~~~ \alpha=2/3,~~~~\gamma =2/3.
  \en
 This fact is very interesting since the equation of state of the perturbed RN-AdS (\ref{qutphiq}) is very different from the unperturbed one (\ref{tphiq}), while their critical behaviors are exactly the same. The result will be especially significant if one is familiar with critical phenomenon of ordinary matters.   For ordinary matter, MFT and RGT present different critical exponents. Theoretically, MFT omits the perturbations around the critical point, while RGT carefully considers the perturbation effects at the critical point. In RGT, the whole system at the critical point is length scale free, that is, there is no special length scale in this system. In a gravity system, there is an inherent length scale $G^{-1/2}$, which makes the RGT cannot do its work in a gravity system. A popular result is that the  $G^{-1/2}$ with a length scale hinders us to renormalize gravity. Here it hinders us to apply RGT in gravity, which makes the perturbed gravity and unperturbed gravity share the same critical exponents, though the perturbation shifts the critical point. This is our main result of this work.

   \section{conclusion}
   In this article, we calculate the critical exponents of RN-AdS with quantum perturbations to all orders. The critical point is shifted due to the perturbations. The perturbations yield a smaller $T_c$, $\phi_c$, and $q_c$ in general. Also, the phase portraits $(T,~\phi,~q)$ of the perturbed RN-AdS and unperturbed RN-AdS seem different. However, the most important parameter to describe the behavior around the critical point, i.e., the critical exponents, are not changed by the perturbation. This is surprising since perturbation is always important at the critical point, as we learn from thermodynamics and statistical mechanics for ordinary matters. A profound reason maybe that there is no complete analogy between thermodynamics and black hole physics. We can easily present some known examples about this gap \cite{Preskill}. Gravity has some special laws, even though we found the close relation between gravitational theory and thermodynamics.

 {\bf Acknowledgments.}
  This work is supported by the Program for Professor of Special Appointment (Eastern Scholar) at Shanghai Institutions of Higher Learning, National Education Foundation of China under grant No. 200931271104, Shanghai Municipal Pujiang grant No. 10PJ1408100, and National Natural Science Foundation of China under Grant No. 11075106.

\end{document}